\documentclass[conference]{IEEEtran}
\IEEEoverridecommandlockouts
\usepackage{cite}
\usepackage{amsmath,amssymb,amsfonts}
\usepackage{algorithmic}
\usepackage{graphicx}
\usepackage{textcomp}
\usepackage{xcolor}
\usepackage{comment}
\def\BibTeX{{\rm B\kern-.05em{\sc i\kern-.025em b}\kern-.08em
    T\kern-.1667em\lower.7ex\hbox{E}\kern-.125emX}}
\begin{document}

\title{Improved Dysarthric Speech to Text Conversion \\ via TTS Personalization \\
\thanks{This work was supported by the Hungarian NRDI Fund 
 through projects NKFIH K143075/K135038, NKFIH-828-2/2021 and by KIFÜ Kommondor.}
}

\author{
\IEEEauthorblockN{
Péter Mihajlik\IEEEauthorrefmark{1}\IEEEauthorrefmark{3},
Éva Székely\IEEEauthorrefmark{2},
Piroska Barta\IEEEauthorrefmark{1},
Máté Soma Kádár\IEEEauthorrefmark{3}\IEEEauthorrefmark{4},
Gergely Dobsinszki\IEEEauthorrefmark{4}\IEEEauthorrefmark{1},
László Tóth\IEEEauthorrefmark{5}
}
\IEEEauthorblockA{\IEEEauthorrefmark{1}Department of Telecommunications and Artificial Intelligence, Budapest University of Technology, Hungary}
\IEEEauthorblockA{\IEEEauthorrefmark{2}Division of Speech, Music and Hearing, KTH Royal Institute of Technology, Sweden}
\IEEEauthorblockA{\IEEEauthorrefmark{3}Hungarian Research Centre for Linguistics, HUN-REN, Hungary}
\IEEEauthorblockA{\IEEEauthorrefmark{4}SpeechTex Ltd., Hungary}
\IEEEauthorblockA{\IEEEauthorrefmark{5}Institute of Informatics, University of Szeged, Hungary}
}
\maketitle

\begin{abstract}
We present a case study on developing a customized speech-to-text system for a Hungarian speaker with severe dysarthria. State-of-the-art automatic speech recognition (ASR) models struggle with zero-shot transcription of dysarthric speech, yielding high error rates. To improve performance with limited real dysarthric data, we fine-tune an ASR model using synthetic speech generated via a personalized text-to-speech (TTS) system. We introduce a method for generating synthetic dysarthric speech with controlled severity by leveraging premorbidity recordings of the given speaker and speaker embedding interpolation, enabling ASR fine-tuning on a continuum of impairments. Fine-tuning on both real and synthetic dysarthric speech reduces the character error rate (CER) from 36–51\% (zero-shot) to 7.3\%. Our monolingual FastConformer\_Hu ASR model significantly outperforms Whisper-turbo when fine-tuned on the same data, and the inclusion of synthetic speech contributes to an 18\% relative CER reduction. These results highlight the potential of personalized ASR systems for improving accessibility for individuals with severe speech impairments.
\end{abstract}


\section{Introduction}
Certain medical conditions can significantly impact an individual's speech production over the long term. For instance, stroke-related cerebral injuries may lead to dysarthric speech, which can be challenging to understand.
Medically, dysarthria is an umbrella term for any neuromotor disorder that results in abnormal speech control. It might influence the speed, strength, accuracy, range, tone, duration etc. of the speech signal, depending on its actual etiology~\cite{duffy2013}. 
Although dysarthric speech often exhibits a significantly altered articulation, leading to low or even near-zero general intelligibility, the speech errors tend to be consistent. As a result, close acquaintances can often learn to understand the altered speech patterns.
Therefore, adapting Automatic Speech Recognition (ASR) systems for the transcription of dysarthric speech is a promising approach, as evidenced by numerous previous studies \cite{bhat2025speech}.
The challenges of dysarthric ASR, however, are not restricted to the disordered speech signal: data collection can be particularly difficult not only because speech production can be slow and tiresome to the speakers affected but also because the human transcription (and its verification) needs special skills and the active involvement of people suffering from dysarthria.  As a result, only a limited amount of dysarthric speech data is available for research and much of it consists of isolated words or expressions ~\cite{kim2008dysarthric}. 

Data scarcity poses a significant challenge, particularly when developing a general speech-to-text system for dysarthric speech, especially in languages with relatively few speakers. Moreover, dysarthria manifests in multiple forms~\cite{bhat2025speech}, making it practically infeasible to design a single ASR system that accommodates all variations.
For assistive technology, personalizing ASR for a single user is a practical approach, as the system is designed solely for their speech patterns~\cite{green2021automatic}.

In this study, we present our results in developing an assistive Hungarian-language speech-to-text system for a person with dysarthria, adapting the ASR system using only a few minutes of dysarthric speech.
To mitigate overfitting in large end-to-end neural acoustic models fine-tuned on limited data, we first develop a personalized text-to-speech (TTS) system to generate dysarthric speech data for augmentation purposes.
Results show that even in this few-shot scenario, the high initial error rate of a foundational ASR model can be significantly reduced. Further improvements are achieved by incorporating TTS-generated speech, synthesized using multiple methods to reflect different levels of dysarthria and intelligibility.
Consequently, we achieved a Character Error Rate (CER) as low as 7.3\% on an independent evaluation set from the same speaker. This suggests that our end-to-end ASR model is already suitable for real-life dysarthric speech transcription and could pave the way for assistive ASR technology for others with similar speech disorders.

\section{Related work}
\label{sec:related_work}

Earlier studies focused on increasing intelligibility~\cite{kain2007improving} -- a direction still active today~\cite{dong2024reconstruction}. As regards ASR, the first attempts aimed at recognizing isolated words only~\cite{fried1985voice,green2003automatic}.
With the advent of the deep neural technologies the the demand for larger amounts of training data has further increased, leading to the ubiquitous use of data augmentation and the utilization synthetic data \cite{xiao2022scaling, yang2024enhancing}.
Hence, many of the current papers on dysarthric ASR focus on how to convert normal speech into 'dysarthric'~\cite{huang2022towards}, or even to synthesize dysarthric speech to use it for the purpose of data augmentation in ASR. 



In the context of dysarthric speech, the benefit of TTS augmentation was shown in~\cite{hermann2023few}, in a few-shot learning scenario. Using FastSpeech 2-based multi-speaker TTS ~\cite{renfastspeech} 
with a learned dysarthria embedding, it was shown that synthetic dysarthric speech improves isolated word recognition of an unseen dysarthric speaker. Notably, while synthetic speech alone was not sufficient to train a model from scratch, combining a small amount of real dysarthric audio with abundant synthetic speech outperformed using only the limited real data. This finding underlines that TTS-generated data can quickly adapt ASR to new dysarthric speakers or commands, especially when paired with even a few authentic samples. 
\cite{soleymanpour2024accurate} applied a multi-speaker TTS approach that injects dysarthria-specific controls into a neural TTS system. Their method adds a dysarthria severity level embedding and a pause-insertion mechanism to an end-to-end TTS, enabling the synthesis of dysarthric speech at various severity levels. Augmenting a dysarthria-tailored ASR system with such synthetic data significantly reduced error rates: a DNN-HMM ASR model trained with the synthetic dysarthric speech achieved a 12.2\% WER reduction over the baseline, and the explicit severity/pause modeling provided an additional 6.5\% WER decrease.
\cite{leung24_interspeech} trained a diffusion-based Grad-TTS \cite{popov2021grad} model from scratch on dysarthric data to create synthetic samples without any parallel typical recordings. Using these samples to fine-tune Whisper led to significant improvements: for the TORGO dysarthric dataset \cite{rudzicz2012torgo}, adding generated data reduced WER from 56.1\% (using only real data) to 20.1\%.

While most augmentation studies aim to improve generic ASR models across a variety of dysarthric speakers \cite{wang2023improving}, an alternative research focus is tailoring ASR to individual users. Personalized dysarthric ASR systems -- trained on a single user’s speech patterns -- have been shown to outperform generalized models on the given user’s speech \cite{green2021automatic} for short phrases, or using small natural dysarthric corpora~\cite{tobin2022personalized}.
Our approach aligns with this personalized paradigm, specifically for assistive technology. By leveraging a pre-dysarthria speech corpus from the user (to capture their voice identity) and adapting a foundation TTS model to produce dysarthric speech, we effectively reproduce the user’s impaired voice for data augmentation. This strategy, inspired by the successes of multi-speaker and fine-tuned TTS in prior work, allows us to bootstrap high-accuracy general speech-to-text conversion with minimal real dysarthric recordings.

\section{Data }
\label{sec:data}
\subsection{Dysarthric Speech Corpora}
\label{ssec:dys-speech-corpora}
Altogether \textbf{less than an hour} of in-domain, Hungarian language transcribed dysarthric speech was available for the research. For the training set, phonetically rich sentences 
were read by the target (stroke survivor) speaker. For the validation set, a short story was read. As for the evaluation, sentences of everyday life were read from the same speaker. Statistics of the dysarthric speech data sets are shown on Table~\ref{tab:dys-speech}.  

\begin{table}[b]
\caption{Single-speaker dysarthric speech data sets}
\begin{center}
\begin{tabular}{|c|c|c|c|}
\hline
& Train-dys & Val-dys & Test-dys \\
\hline
Duration [min] & 21 & 6 & 17 \\ 
\hline
\# segments & 195 & 40 & 107\\
\# words & 1406 & 363 & 1382\\
\# chars & 9523 & 2331 & 8326\\
\hline
\end{tabular}
\label{tab:dys-speech}
\end{center}
\end{table}

\subsection{Fluent Speech Corpora}
\label{ssec:fluent-speech-corpora}
In our particular case, the target speaker suffering from dysarthria recorded \textbf{13 hours} of spontaneous speech in \textbf{lecture materials} (as part of his profession as a university teacher) prior to the stroke event. Although such amount of healthy speech might not be typical for people developing dysarthria in general, due to the wide spread use of social media, similar scenario might become more common in the future. This collection of lecture speech is, however, unlabeled.

Additionally, the same \textbf{195 phonetically rich sentences} read in the Train-dys set were recorded previously (before stroke) as well by the target speaker. This is the only supervised, i.e., exactly transcribed, fluent speech data set from the given subject.

\subsection{Synthetic Dysarthric Speech Corpora}
\label{ssec:syn-dys-speech-corpora}
As the amount of in-domain (dysarthric) speech was minuscule -- even for speaker-specific ASR -- and because of the domain mismatch regarding the fluent (unlabeled) speech from the same speaker, we have decided to create more dysarthric data by speech synthesis utilizing the previous data sources and a foundational TTS model.
The  TTS system XTTS-v2 \cite{casanova24_interspeech} was selected for the generation of dysarthric Hungarian speech for the target speaker, because it is an open-sourced multilingual deep neural network based model that includes support for the Hungarian language\footnote{https://huggingface.co/coqui/XTTS-v2}. 

\subsubsection{Adaptation to the target speaker's original voice}
The corpus of 13 hours of lectures was segmented into utterances delineated by breath events following \cite{szekely2019casting}.
To select the utterances from the lecture recordings which are most suitable for TTS training (regarding fluency, intelligibility) we employed an ASR ensemble technique using automatic transcriptions created with both BEAST2 \cite{beast2} and Whisper \cite{radford2023robust}.
A subset of the lecture corpus was created by comparing the two ASR transcriptions and selecting the utterances where the two transcriptions had an edit-distance of max 4 characters.  
This yielded 1242 utterances, corresponding to \textbf{2 hours of lecture speech} associated with automatic transcriptions (pseudo-labels). 

\subsubsection{Fine-tuning on dysarthric speech}
\label{sssec:dysarthricTTS}
To approximate the increased variability of dysarthric speech resulting from speakers articulating words with varying degrees of success, we leverage premorbidity recordings of the same speaker to generate synthetic dysarthric speech at different severity levels. Since these recordings provide a clean baseline of the speaker’s healthy voice, we fine-tune the speaker-adapted TTS model on the 195 recorded sentences of dysarthric speech (Train-dys) to gradually shift the synthesis away from fluent articulation toward impaired speech. By controlling the extent of fine-tuning and interpolating between fluent and dysarthric embeddings, we aim to create a continuum of severity that better reflects the fluctuating nature of dysarthric speech.
To enable synthesis with different severity levels of dysarthria, we saved 3 checkpoints:

\begin{itemize}
\item \textit{FT-Dys-Undertrained} (training stopped after 2000 iterations, while validation loss was still decreasing)

\item \textit{FT-Dys-Best} (optimally trained, validation loss minimised after 3200 iterations)

\item \textit{FT-Dys-Overtrained} (trained for 3000 iterations after validation loss minimised; overfitted to dysarthric speech)
\end{itemize}

Additionally, the speaker-adapted XTTS model was fine-tuned on the 195 recorded sentences of both dysarthric speech and on the 195 recorded sentences of typical speech using 2 separate speaker embeddings until the validation loss was minimized; we call this model FT-Dys-Co-trained. The goal of this approach was to be able to scale the level of dysarthria by weighting the two embeddings.

\subsubsection{Synthesizing dysarthric Speech}
\label{ssec:synthetic_data}
We carefully selected the text material for synthesis:
5000 sentences were filtered from the spoken language subset of the Hungarian Gigaword corpus \cite{oravecz-etal-2014-hungarian} with lengths of 9--11 words not containing foreign words or abbreviations.

We synthesized these sentences with \textit{FT-Dys-Undertrained, FT-Dys-Best, FT-Dys-Overtrained}, using temperature = 0.9, repetition-penalty = 6 parameters. To ensure maximum variability, the 195 dysarthric sentences in the training data were looped through as reference audio so each of them was used 12-13 times. 
\cite{casanova24_interspeech} reports that both speaker and style characteristics of the reference audio are copied during synthesis.

\textit{FT-Dys-Co-trained} was used to synthesize 4 additional datasets. We modified the interpolation function of XTTS -- which originally performed an equal-weight blend of speaker embeddings --, to support explicit weighted combinations. The new function accepts arbitrary weight vectors -- when the weights sum to one, they yield a convex combination that precisely controls the relative contributions of each speaker embedding. Moreover, by permitting negative weight values, the function facilitates \textit{extrapolation} in the speaker embedding space, thereby enabling modulation of speaker characteristics beyond the range defined by the reference embeddings.
Using the same looping method for reference audios from each corpus [typical, dysarthric], we synthesized the 5000 sentences with settings A:[0.2, 0.8], B:[0.0, 1.0], C:[-0.5, 1.5], D:[-1.5, 2.5], (temperature=0.9, repetition-penalty=6).
The total duration of the resulted synthesized data sets varied between 10 and 12 hours.


\section{Zero-Shot ASR Experiments}
\subsection{Comparison of ASR Engines on Dysarthric Speech}
The aim of these experiments was to assess the difficulty of automatic transcription of dysarthric speech in Hungarian from the target speaker by comparing various ASR engines. As Table~\ref{tab:zs-asr-dys-speech} shows, state-of-the-art multilingual and monolingual deep neural approaches fail to deliver a transcription with any practical usability. The last one 'FastConformer\_Hu', is our in-house ASR model \cite{Dobsinszki2025} trained on 5k hours of Hungarian broadcast and conversational speech  following the 'FastConformer-BPE-CTC' recipe of the NVIDIA NeMo toolkit \cite{nemo} and using the FastConformer architecture \cite{rekesh2023fast}.  As it  consistently achieved the lowest error rates, we focused on the FastConformer\_Hu ASR model (115 Million parameters), but control experiments were conducted with the Whisper-large-v3-turbo model (Whisper-turbo, in short, with 809 Million parameters), as well. 

\begin{table}[t!]
\caption{Zero-shot (baseline) ASR results [\%] on single speaker Hungarian dysarthric speech data sets}
\begin{center}
\begin{tabular}{|l|cc|cc|cc|}
\hline
& \multicolumn{2}{c|}{\textbf{Train-dys}} & \multicolumn{2}{c|}{\textbf{Val-dys}} & \multicolumn{2}{c|}{\textbf{Eval-dys}} \\
\textbf{ASR engine} & \textbf{WER} & \textbf{CER} & \textbf{WER} & \textbf{CER} & \textbf{WER} & \textbf{CER} \\
\hline
Gemini 2.0 Flash\footnotemark & 101 & 66 & 74 & 45 & 76 & 51 \\ 
Whisper-turbo\cite{radford2023robust} & 107 & 69 & 62 & 34 & 68 & 44 \\ 
BEAST2\cite{beast2} & 76 & 44 & 68 & 38 & 68 & 43 \\ 
\hline
FastConformer\_Hu & 72 & 36 & 58 & 22 & 65 & 36 \\ 
\hline
\end{tabular}
\label{tab:zs-asr-dys-speech}
\end{center}
\end{table}
\footnotetext{https://deepmind.google/technologies/gemini/flash/}

\subsection{Intelligibility-related ASR tests on Synthesized Data}

Using a well-trained general ASR model for re-transcribing synthesized speech and evaluating WER/CER is a common practice to estimate the intelligibility of various TTS approaches. Based on the previous results, we have decided to apply the FastConformer\_Hu model for this purpose. In Table~\ref{tab:zs-asr-synth-dys-speech} it can be clearly seen that the under- and overtrained XTTS models gave lower and higher WER/CER, respectively, as expected. By comparing the Co-trained A-D approaches again, a monotonic increase (decreasing intelligibility) can again be observed. To verify the appropriateness of the FastConformer\_Hu ASR model, typical (premorbidity) speech data sets from the target speaker were additionally transcribed and evaluated, confirming that fluent read and lecture speech is the easiest to transcribe (and understand). Based on the WER/CER, we can conclude that the various 
synthetic corpora resemble different severity levels of dysarthric speech.
However, as expected with fine-tuning approaches, the WER/CER of the natural dysarthric speech is still higher (Train-dys, see Table~\ref{tab:zs-asr-dys-speech}).

\begin{table}[h!]
\caption{Zero-shot FastConformer\_Hu ASR results [\%] on synthesized dysarthric speech \\ and on healthy control speech}
\begin{center}
\begin{tabular}{|l|cc|}
\hline
\textbf{Data set} & \textbf{WER} & \textbf{CER}  \\
\hline
FT-Dys-Undertrained  & 44 & 16  \\ 
FT-Dys-Best  & 46 & 17  \\
FT-Dys-Overtrained  & 60 & 26  \\
\hline
FT-Dys-Co-trained-A  & 35 & 12  \\
FT-Dys-Co-trained-B  & 45 & 17  \\
FT-Dys-Co-trained-C  & 54 & 23  \\
FT-Dys-Co-trained-D  & 57 & 26  \\
\hline\hline
Fluent phon. rich sentences  & 4.1 & 0.3  \\
2 hours of lectures\footnotemark & 9.8 & 2.4  \\
\hline
\end{tabular}
\label{tab:zs-asr-synth-dys-speech}
\end{center}
\end{table}
\footnotetext{Pseudo-labels were used as references.}

\section{ASR Model Fine-tuning on Dysarthric Speech}

In this set of experiments the FastConformer\_Hu and Whisper-turbo ASR models used in the previous (zero-shot) experiments were fine-tuned on various dysarthric speech data sets. The Fastconformer\_Hu was fine-tuned in the NeMo environment \cite{nemo}, keeping the original recipe by default. The learning-rate was optimized for each setup, and the following hyperparameters were kept constant: batch size=16, number of iterations=16k, AdamW optimizer with betas [0.8, 0.9], weight decay=0.02, cosine annealing, minimum learning rate=1e-7. In case of Whisper-turbo we used the HuggingFace Transformer library and recipe\footnote{https://huggingface.co/blog/fine-tune-whisper}. Similarly to the other ASR model, learning rate was optimized for each setup and we fixed the maximum number of steps to 5000 for fine-tuning. All experiments were run on one RTX 5000 Ada GPU. 

We found that leaving the original dysarthric train set (Train-dys) out always deteriorated the results, so, by default, it was added to the fine-tune sets. Beyond investigating the effect of fine-tuning on each $\sim$10 hours long synthetic dysarthric speech data set (along with Train-dys), we also applied combinations of synthetic sets (Best + Overtrained vs. all of them). In the experiments, by default, the ratio of original/synthesized speech (21 min/10 hours) was kept constant. Finally we conducted control experiments with mixed fine-tune sets (21 min. from Train-dys + 2 hours of fluent lectures) and by fine-tuning  on fluent speech only. For results of FastConformer\_Hu  and Whisper-turbo, see Table~\ref{tab:ft-asr-synth-dys-speech} and Table~\ref{tab:ft-whisper-synth-dys-speech}, respectively.

\begin{table}[t]
\caption{Fine-tuned \textbf{FastConformer\_Hu} ASR results [\%] \\ measured on dysarthric speech}
\begin{center}
\begin{tabular}{|l|cc|cc|}
\hline
& \multicolumn{2}{c|}{\textbf{Val-dys}} & \multicolumn{2}{c|}{\textbf{Eval-dys}} \\
\textbf{Fine-tuning data sets} & \textbf{WER} & \textbf{CER}  & \textbf{WER} & \textbf{CER}  \\
\hline
Train-dys  & 28.4 & 9.2  & 21.9 & 8.9  \\ 
\hline
Train-dys + FT-Dys-Undertrained & 24.0 & 8.1  & 22.5 & 9.2  \\
Train-dys + FT-Dys-Best & 22.8 & 7.7  & 19.5 & 7.7  \\
Train-dys + FT-Dys-Overtrained  & \textbf{22.6} & \textbf{7.4}  & 19.0 & 7.8  \\
\hline
Train-dys + FT-Dys-Co-trained-A  & 24.0 & 7.8  & 21.0 & 8.8  \\
Train-dys + FT-Dys-Co-trained-B  & 24.2 & 8.3  & 22.7 & 9.4  \\
Train-dys + FT-Dys-Co-trained-C  & 24.3 & 8.9  & 22.7 & 9.5  \\
Train-dys + FT-Dys-Co-trained-D  & 25.9 & 8.9  & 23.3 & 10.0  \\
\hline
Train-dys + FT-Dys-Best + Over  & 26.7 & 8.6 & \textbf{18.3} &\textbf{7.3}  \\
Train-dys + All (7) FT-Dys & 25.1 & 9.0 & 19.7 & 8.3  \\
\hline\hline
Train-dys + 2 hours of lectures & 27.6 & 9.1 & 26.0 & 10.7  \\
Fluent phon. rich sentences & 54.8 & 22.3 & 59.3 & 31.8  \\
\hline
\end{tabular}
\label{tab:ft-asr-synth-dys-speech}
\end{center}
\end{table}

\begin{table}[b]
\caption{Fine-tuned \textbf{Whisper-turbo} ASR results [\%] \\ measured on dysarthric speech}
\begin{center}
\begin{tabular}{|l|cc|cc|}
\hline
& \multicolumn{2}{c|}{\textbf{Val-dys}} & \multicolumn{2}{c|}{\textbf{Eval-dys}} \\
\textbf{Fine-tuning data sets} & \textbf{WER} & \textbf{CER}  & \textbf{WER} & \textbf{CER}  \\
\hline
Train-dys  & 41.1 & 14.9  & 35.0 & 14.7  \\ 
\hline
Train-dys + FT-Dys-Undertrained & 36.6 & 13.3  & 35.2 & 15.2  \\
Train-dys + FT-Dys-Best & 35.8 & 12.1 & 32.2 & 13.1  \\
Train-dys + FT-Dys-Overtrained  & 35.6 & 11.8  & 31.8 & 13.1  \\
\hline
Train-dys + FT-Dys-Co-trained-A  & 35.5 & 11.2 & 33.4 & 14.1  \\
Train-dys + FT-Dys-Co-trained-B  & \textbf{33.9} & 11.4  & 34.7 & 14.2  \\
Train-dys + FT-Dys-Co-trained-C  & 36.9 & 11.8  & 34.0 & 14.0  \\
Train-dys + FT-Dys-Co-trained-D  & 35.5 & \textbf{10.9}  & 35.5 & 14.6  \\
\hline
Train-dys + FT-Dys-Best + Over  & \textbf{33.9} & 11.4 & 32.0 & 12.3  \\
Train-dys + All (7) FT-Dys & 34.7 & \textbf{10.9} & \textbf{30.4} & \textbf{12.2}  \\
\hline
\end{tabular}
\label{tab:ft-whisper-synth-dys-speech}
\end{center}
\end{table}

\section{Results and Discussion}
\begin{figure}[t]
    \centering
    \includegraphics[width=\columnwidth]{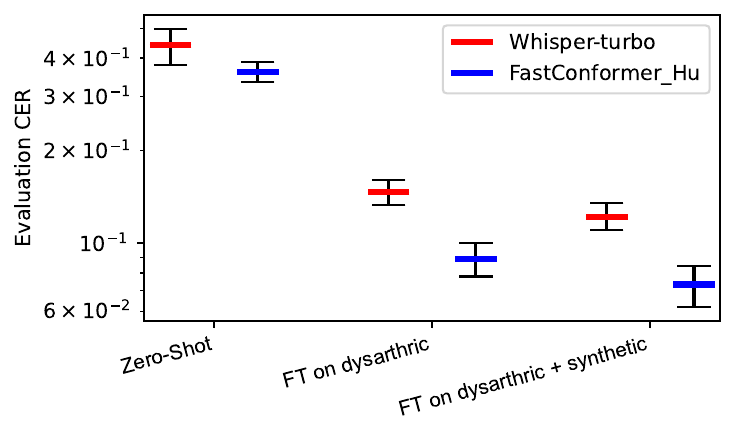}
    \caption{Best CER results on dysarthric speech (Eval-dys) for different ASR approaches: zero-shot, fine-tuning (FT) only on real dysarthric speech, and FT on real + synthesized dysarthric data (with confidence intervals).}
    \label{fig:confidence_intervals}
\end{figure}

As the results
show, ASR models fine-tuned on dysarthric speech gave dramatically lower error rates on both the in-domain validation and evaluation sets containing speech with severe dysarthria than the baseline systems (Table~\ref{tab:zs-asr-dys-speech}). Moreover, many of the synthesized dysarthric corpora further reduced the error rates on the validation, evaluation sets, or both. Considering our most important metric, evaluation CER, however, the best results 
were achieved when combining synthetic dysarthric data sets in addition to real dysarthric speech. We focus on CER rather than WER, as it better reflects the manual effort required to correct ASR output and provides a more stable statistical measure.
The improvements from combining synthetic dysarthric data with real dysarthric fine tuning are evident for both the FastConformer\_Hu and Whisper-turbo models. As Figure~\ref{fig:confidence_intervals} shows, the best results are outside of the confidence intervals (calculated using the tool of Ferrer$\&$Riera~\cite{Confidence_Intervals}) of the 'train-dys' only fine-tuning setup, confirming the effectivity of our approach.


Comparing the foundational ASR models FastConformer\_Hu and Whisper-turbo, the conclusions are straightforward: while Whisper offers high accuracy and accessibility for English, our well-trained monolingual model could significantly outperform it in the given Hungarian language task. 
Finally, the control experiment results  (the last two rows of Table~\ref{tab:ft-asr-synth-dys-speech}) show that adding typical (premorbidity) data from the target speaker, instead of synthetic dysarthric data, did not generally improve the results. 
This confirms that the observed improvements were not due to general speaker-specific characteristics (such as vocal tract length or timbre) but rather to the effectiveness of few-shot learning in modeling dysarthric articulation.
The high error rates from fine-tuning only on typical (and supervised) speech indicate that adapting to typical speech has little impact 
on recognizing dysarthric speech.


\section{Conclusion}
This study demonstrates that a personalized ASR system designed for assistive technology can enable accurate speech-to-text conversion for individuals with severe dysarthria, even with minimal dysarthric speech data in a relatively low-resource language. By fine-tuning a monolingual FastConformer-based ASR model and augmenting it with synthetic dysarthric speech, we reduced the character error rate (CER) from 36–51\% (zero-shot) to 7.3\% and the word error rate (WER) from 76–65\% to 18.3\%. The monolingual model outperformed Whisper-turbo, highlighting the importance of language-specific adaptation.
We used a speaker-adapted TTS system trained on premorbidity recordings to generate dysarthric speech with controlled severity. Synthetic data improved ASR performance, contributing to an 18\% relative CER reduction. Control experiments confirmed that adaptation to fluent speech alone did not enhance dysarthric ASR, reinforcing the need for targeted modeling of impaired articulation.
Given the promising results of zero-shot TTS which can utilize as little as a few seconds of unaffected recordings from a speaker \cite{casanova24_interspeech}, our results are likely reproducible with much less typical speech data than the what was used in this study, which will be subject of future work.

\bibliographystyle{IEEEtran}  
\bibliography{references}  

\end{document}